\documentclass[preprint,aps,prd]{revtex4-1}
\usepackage{graphicx}
\usepackage{slashed}
\usepackage[colorlinks]{hyperref}
\usepackage{cancel,bm} 
\usepackage{amsmath,amssymb}
\usepackage{mathtools}
\usepackage{float}
\usepackage{color}
\usepackage{leftidx}
\usepackage{booktabs}
\usepackage{latexsym}
\usepackage{revsymb}
\usepackage{multirow}
\usepackage{hypcap}
\usepackage{array}

\hypersetup{
    colorlinks=red,
    citecolor=blue,
    filecolor=magenta,      
    urlcolor=blue,
}

\begin{document}

\title{Precision neutrino data confronts $\mu\leftrightarrow\tau$ symmetry}

\author{Rambabu Korrapati}

\email{rambabu@phy.iitb.ac.in}.
    
\affiliation{ Department of Physics,
Indian Institute of Technology Bombay, Mumbai-400076,
India.}

\author{S. Uma Sankar}

\email{uma@phy.iitb.ac.in}.
    
\affiliation{ Department of Physics,
Indian Institute of Technology Bombay, Mumbai-400076,
India.}

\date{\today}

\begin{abstract}
Neutrino oscillation data indicate that $\theta_{23}$ is close to $\pi/4$ and $\theta_{13}$
is very small. A simple $\mu\leftrightarrow\tau$ exchange symmetry of the neutrino mass matrix
predicts $\theta_{23}=-\pi/4$ and
$\theta_{13}=0$.
Since the experimental measurements differ from these predictions,
this symmetry is obviously broken. This breaking is given by two parameters: $\varepsilon_1$
parametrizing the inequality bewteen $12$ and $13$ elements and $\varepsilon_2$
parametrizing the inequality bewteen $22$ and $33$ elements. We show that the magnitude
of $\theta_{13}$ is essentially controlled by $\varepsilon_1$ whereas the deviation
of $\theta_{23}$ from maximality is controlled by $\varepsilon_2$. The measured value of $\theta_{13}$ requires 
$\mu\leftrightarrow\tau$ symmetry to be badly broken for both normal hierarchy
and inverted hierarchy, though the level of breaking depends sensitively on the 
hierarchy. In this paper
we obtain constraints on  the parameters of neutrino mass matrix, including the
symmetry breaking parameters, using the precision
oscillation data. We find that this precision data constrains all elements
of neutrino mass matrix to be in very narrow ranges. We also consider
$\mu\leftrightarrow -\tau$ exchange symmetry in the case of inverted hierarchy
and find that it provides an explanation
of neutrino mixing angles with some fine-tuning.
\end{abstract}

\maketitle
\raggedbottom 

\section{Introduction}
The data from solar \cite{Cleveland:1998nv, Fukuda:1996sz, Abdurashitov:1994bc, Hampel:1998xg, Altmann:2005ix, Abe:2016nxk, Bahcall:2001zu}
and atmospheric \cite{Casper:1990ac, Nakahata:1986zp, Ashie:2005ik}  
neutrino experiments have        
provided a strong hint of neutrino oscillations. Later experiments
with man made sources measured the neutrino oscillation parameters
precisely. These precision measurements lead to stringent constraints on the elements
of neutrino mass matrix.

The three flavor states $\nu_\alpha$ ($\alpha=e,\mu,\tau$) mix among
themselves to form three mass eigenstates $\nu_i$ ($i=1,2,3$) which have
well-defined mass eigenvalues $m_1,m_2$ and $m_3$. The flavor eigenstates
are related to the mass eigenstates through the Pontecorvo-Maki-Nakagawa-Sakata (PMNS) mixing
matrix $U$ \cite{Pontecorvo:1967fh, Maki:1962mu} as
\begin{equation}
 \nu_\alpha=\sum_iU_{\alpha i}\nu_i.
\end{equation}
The elements $U_{\alpha i}$ depend on three mixing angles, $\theta_{12},
\theta_{13}$ and $\theta_{23}$ and the CP violating phase ($\delta_{\rm CP}$). 
From the three mass eigenvalues we can define three mass-squared differences
$\Delta m^2_{ij}=m^2_i-m^2_j$, of which only two are independent.
It is known that the mass-squared difference needed to solve the
solar neutrino anomaly is much smaller than that to solve the atmospheric
neutrino anomaly. Hence we choose $\Delta m^2_{21}$ to be the smaller
mass-squared difference, which we label as $\delta m^2$ and $\Delta m^2_{31}$
to be the larger mass-squared difference.
The third mass-squared difference, $\Delta m^2_{32}=\Delta m^2_{31}-\Delta m^2_{21}$,
is approximately equal to $\Delta m^2_{31}$. We define the average of
$\Delta m^2_{31}$ and $\Delta m^2_{32}$ to $\Delta m^2$. The neutrino oscillation
probabilities depend on the two independent mass-squared differences,
$\delta m^2$ and $\Delta m^2$, the three mixing angles $\theta_{ij}$
and the $\delta_{\rm CP}$ phase.

 The expression
for the most general three flavor oscillation probability is
\begin{eqnarray}\label{3fprob}
 P(\nu_\alpha\to \nu_\beta)&=&\delta_{\alpha\beta}-4\sum_{i>j}{\rm Re}(U^*_{\alpha i}U_{\beta i}U_{\alpha j}U^*_{\beta j})
 \sin^2\left(1.27\frac{\Delta m^2_{ij}L}{E}\right)       \nonumber   \\
&& -2\sum_{i>j}{\rm Im}(U^*_{\alpha i}U_{\beta i}U_{\alpha j}U^*_{\beta j})\sin\left(2.54\frac{\Delta m^2_{ij}L}{E}
\right).
\end{eqnarray}
In principle it is a difficult procedure to determine the oscillation parameters
from any experiment given the complicated expression in eq.~(\ref{3fprob}).
However  two of the parameters in neutrino oscillation formalism are small. 
CHOOZ experiment set the upper limit 
$\sin^22\theta_{13}\leq 0.1$, implying that $\theta_{13}$ is small.
Solar and atmospheric data show that the ratio $\delta m^2/\Delta m^2 \ll 1$.
The smallness of these two quantities enable us to make precision measurements
of the mass-squared differences and the mixing angles.

For the long baseline reactor experiment KamLAND \cite{Gando:2010aa}, we have $L\sim 180$ km    
and $E\sim 5$ MeV. For these values we find that $\delta m^2 L/E\sim 1$ and
$\Delta m^2L/E\gg 1$. If we substitute $\theta_{13}=0$ in the expression
for survival probability of electron anti-neutrinos, we get
\begin{equation}
P(\bar{\nu}_e\to\bar{\nu}_e)=1-\sin^22\theta_{12}\sin^2\left(1.27\frac{\delta m^2L}{E}\right).
\end{equation}
In the approximation of neglecting small $\theta_{13}$, we find that the
data of KamLAND experiment can be interpreted in terms of an effective
two flavor oscillation formula governed by $\delta m^2$ and $\theta_{12}$.
The spectral distortion
data of KamLAND \cite{Araki:2004mb} leads to a very precise determination of $\delta m^2$         
and a moderately precise determination of $\tan^2\theta_{12}$:
\begin{equation}
 \delta m^2 = 7.9^{+0.6}_{-0.5}\times10^{-5}\ {\rm eV^2}\ {\rm and}\ \tan^2\theta_{12}=0.4^{+0.10}_{-0.07}.
\end{equation}
Solar neutrino data requires $\delta m^2$ data to be positive.
For long baseline accelerator experiment MINOS \cite{Adamson:2013whj} we have $L\sim 730$ km and  
$E \sim 3$ GeV. For these values we find that $\delta m^2 L/E\ll 1$ and
$\Delta m^2L/E\sim 1$. Hence we set $\delta m^2$ and $\theta_{13}$
both equal to zero in the expression for the survival probability
of the muon neutrinos. This leads to 
\begin{equation}
P(\nu_\mu\to\nu_\mu)=1-\sin^22\theta_{23}\sin^2\left(1.27\frac{\Delta m^2L}{E}\right).
\end{equation}
Once again we have an effective two flavor formula. Analyzing the data of MINOS
with this formula leads to precise values of $|\Delta m^2|$ and $\sin^22\theta_{23}$:
\begin{equation}
 |\Delta m^2| = 2.41^{+0.09}_{-0.10}\times10^{-3}\ {\rm eV^2}\ {\rm and}\ \sin^22\theta_{23}=0.950^{+0.035}_{-0.036}.
\end{equation}
For short baseline reactor experiments  Double-CHOOZ \cite{Abe:2011fz},                    
Daya-Bay \cite{An:2012eh} and RENO                                                         
\cite{Ahn:2012nd} we have $L\sim 1$ km and $E\sim 5$ MeV.                                  
 If we substitute $\delta m^2=0$ in the expression
for survival probability of electron anti-neutrinos for these experiments,
we again get the effective two flavor expression
\begin{equation}
P(\bar{\nu}_e\to\bar{\nu}_e)=1-\sin^22\theta_{13}\sin^2\left(1.27\frac{\Delta m^2L}{E}\right).
\end{equation}
Using the value of $\Delta m^2$ from MINOS
experiment, the value $\sin^2\theta_{13}$ is measured to be \cite{Huang:2016gkb}           
\begin{equation}
 \sin^22\theta_{13} = 0.0841\pm0.0027{\rm (stat.)}\pm0.0019{\rm (syst.)}.
\end{equation}

In Table~1 we have shown the results of the global analysis of 
all neutrino oscillation data, including solar,
atmospheric, reactor and accelerator sources \cite{Esteban:2016qun}.              
\begin{table}[htbp]
\centering
  \begin{tabular}{|c|c|c|c|}
\hline
    Parameter                                                        & Best Fit   & 1$\sigma\ {\rm range}$             & 3$\sigma\ {\rm range}$   \\ 
\hline
   $\delta m^2/10^{-5} {\rm\ eV^2\ (NH\ or\ IH)}$                    & 7.50       & 7.33 - 7.69                        & 7.03 - 8.09              \\ 
\hline
    $\sin^2\theta_{12} {\rm\ (NH\ or\ IH)}$                          & 0.306      & 0.294 - 0.318                      & 0.271 - 0.345            \\
\hline
   $\Delta m^2/10^{-3} {\rm\ eV^2\ (NH)}$                            & 2.524      & 2.484 - 2.563                      & 2.407 - 2.643            \\
   $\Delta m^2/10^{-3} {\rm\ eV^2\ (IH)}$                            & -2.514     & -2.555 - -2.476                    & -2.635 - -2.399          \\
\hline
   $\sin^2\theta_{13} {\rm\ (NH)}$                                   & 0.02166    & 0.02091 - 0.02241                  & 0.01934 - 0.02392        \\
   $\sin^2\theta_{13} {\rm\ (IH)}$                                   & 0.02179    & 0.02103 - 0.02255                  & 0.01953 - 0.02408        \\
\hline
   $\sin^2\theta_{23} {\rm\ (NH)}$                                   & 0.441      & 0.420 - 0.468                      & 0.385 - 0.635            \\
   $\sin^2\theta_{23} {\rm\ (IH)}$                                   & 0.587      & 0.563 - 0.607                      & 0.393 - 0.640            \\
\hline\hline
  \end{tabular}
  \caption{Global Data of three neutrino mass-mixing parameters \cite{Esteban:2016qun}}
\label{globalneutrinodatatable}
\end{table}

From this data, we note the following features: 
\begin{itemize}
 \item Neutrino oscillation data does not give any information on
the lowest value of neutrino mass. It can be almost zero or be equal
to the upper limit from Tritium beta decay of $0.2$ eV \cite{Kraus:2004zw}.
 \item Since the sign of $\Delta m^2_{31}$ is not known, we need to
consider both possible signs. For $\Delta m^2_{31}$ positive, called
the normal hierarchy (NH), the lowest
mass is $m_1$ and the highest mass is $m_3$.
For $\Delta m^2_{31}$ negative, called the inverted hierarchy (IH),
the lowest mass is $m_3$ and the highest mass is $m_2$.
 \item The neutrino mass eigenstates $\nu_i$ ($i=1,2,3$) are identified
by their $\nu_e$ flavor content, which is largest for $\nu_1$ and smallest
for $\nu_3$.
 \item Among the mixing angles, $\theta_{23}$ is close to maximal and 
$\theta_{13}$ is quite small.
\end{itemize}

Various discrete symmetries of the neutrino mass matrix have been proposed to account for the patterns
observed in neutrino masses and mixing angles. The simplest of these
is the $\mu\leftrightarrow\tau$ exchange symmetry of neutrino mass matrix \cite{Harrison:2002et}. 
 This symmetry predicts $\theta_{23}=-\pi/4$ and $\theta_{13}=0$. In this
paper, we will study 
\begin{itemize}
 \item the pattern of $\mu\leftrightarrow\tau$
symmetry breaking to obtain viable values of $\theta_{13}$ and $\theta_{23}$ and
 \item the constraints imposed on the parameters of
neutrino mass matrix by the precision oscillation data.
\end{itemize}

\section{$\mu\leftrightarrow\tau$ Symmetry}
We assume neutrinos are Majorana fermions and the light neutrino mass
matrix is generated through a see-saw mechanism.
The Majorana mass matrix for light neutrinos is a complex symmetric matrix. In this work we assume
it to be real, which (a) simplifies the discussion and (b) makes the analysis more predictive:
\begin{equation}\label{symmetric}
 M_0 = \left(\begin{array}{ccc}
 M_{ee} & M_{e\mu} & M_{e\tau}     \\
M_{e\mu} & M_{\mu\mu} & M_{\mu\tau} \\
M_{e\tau} & M_{\mu\tau} & M_{\tau\tau}        
             \end{array}\right)
= \left( \begin{array}{ccc}
a   & b_1 & b_2 \\
b_1 & c_1 & d \\
b_2 & d   & c_2 \end{array} \right).
\end{equation}
Imposing the $\mu\leftrightarrow\tau$  symmetry \cite{Fukuyama:1997ky, Ma:2001mr, Lam:2001fb, Xing:2015fdg} 
on this mass matrix leads to $b_1 = b_2 = b$
and $c_1 = c_2 = c$. This real symmetric matrix is diagonalized by the orthogonal matrix,
\begin{equation}
 \left( \begin{array}{ccc}
\cos\theta_{12} &  \sin\theta_{12} & 0  \\
-\frac{1}{\sqrt{2}}\sin\theta_{12} & \frac{1}{\sqrt{2}}\cos\theta_{12} & -\frac{1}{\sqrt{2}}  \\
-\frac{1}{\sqrt{2}}\sin\theta_{12}  &  \frac{1}{\sqrt{2}}\cos\theta_{12}   &  \frac{1}{\sqrt{2}} 
\end{array} \right).
\end{equation}
By inspection we can identify $\theta_{13} = 0$ and $\theta_{23} = -\pi/4$ and the value of
$\theta_{12}$ is given by
\begin{equation}\label{th12}
 \tan2\theta_{12} = \frac{2\sqrt{2}b}{c+d-a}.
\end{equation}
The mass eigenvalues are given by
\begin{eqnarray}\label{eigenvalues}
m_1 &=& \frac{a+c+d-k}{2}      \nonumber \\ 
m_2 &=& \frac{a+c+d+k}{2}    \nonumber \\
m_3 &=& c-d,
\end{eqnarray}
where $k = \sqrt{(c+d-a)^2+8b^2}$.
The measured value of $\theta_{12}$ leads to $\sin^2\theta_{12} \approx 1/3$. Substituting
it in the above equation leads to the two relations
\begin{equation}\label{sinetheta12}
 b = c + d - a \ {\rm and} \ k = 3b.
\end{equation}

The expressions for the mass-squared differences are obtained to be
\begin{eqnarray}\label{small:m}
\delta m^2 &=& m^2_2 - m^2_1 \nonumber \\
&=& k(a+c+d)
\end{eqnarray} and
\begin{eqnarray}\label{big:M}
\Delta m^2 &=& m^2_3 - \frac{m^2_1 + m^2_2}{2} \nonumber \\
 &=& \frac{1}{2}\left[(c-d)^2 - 4cd - a^2 -4b^2\right].
\end{eqnarray}
Since only the magnitude of
$\Delta m^2$ is measured there is a sign ambiguity in the constraint of eq.~(\ref{big:M}).
All the four parameters of the
neutrino mass matrix can be exactly determined provided (a) this sign ambiguity is resolved
and (b) the lowest mass eigenvalue is known. In the following, we take 
the lowest mass eigenvalue to be negligibly small. With this assumption, we will
work out the values for neutrino mass eigenvalues and the neutrino mass matrix parameters
for the two cases of normal hierarchy (NH, $m_3 > m_2 \gg m_1$) and inverted hierarchy (IH, $m_2 \geq m_1 \gg m_3$).

\subsection{Normal Hierarchy}

For normal hierarchy, $\Delta m^2$ is positive and we choose $m_1$ to be
negligibly small. This assumption leads
\begin{equation}
 a + c + d \approx k = 3b \ {\rm and} \ \delta m^2 \approx k^2, 
\end{equation}
yielding
\begin{equation}\label{defb}
 b = \frac{k}{3} \approx \frac{\sqrt{\delta m^2}}{3}.
\end{equation}
Combining with the condition from eq.~(\ref{sinetheta12}), we get
\begin{equation}
 a \approx b \ {\rm and} \ c + d \approx 2b.
\end{equation}
From the expression of $\Delta m^2$ in eq.~(\ref{big:M}), we note that
\begin{equation}
 \Delta m^2 \approx \frac{1}{2}[(c - d)^2 - 4cd], 
\end{equation}
which is satisfied if 
\begin{equation}\label{abcdnh}
 c \approx -d \approx \frac{\sqrt{\Delta m^2}}{2} \gg a, b.
\end{equation}
From eqs.~(\ref{th12}) and~(\ref{abcdnh}), we see that the large value
of $\theta_{12}$ arises due to a fine-tuned cancellation in
$(c+d-a)$, which makes its value equal to $b$. From eqs.~(\ref{defb})
and~(\ref{abcdnh}), we see that this cancellation is of the order
$\sqrt{\delta m^2/\Delta m^2}$.
Thus the four parameters of the neutrino mass matrix are determined
exactly by the four conditions, given by 
the three measured parameters $\sin^2\theta_{12}$,
$\delta m^2$ and $\Delta m^2$ and the assumption on the lowest
mass eigenvalue.

We impose the less rigid constraint that the measured values should be
within their $3\sigma$ ranges, as given below
\begin{align}\label{NHunpertconstraints}
0.271 \leq \sin^2\theta_{12} \leq 0.345                    \nonumber  \\
7.03\times10^{-5} \leq \delta m^2 \leq 8.09\times10^{-5}   \nonumber  \\
2.407\times10^{-3} \leq \Delta m^2 \leq 2.643\times10^{-3}    \nonumber  \\
|m_1| < 0.1 \ m_2.
\end{align}
The allowed ranges of the $a, b, c$ and $d$ are
\begin{eqnarray}\label{string}
a &=& 0.0017 - 0.0036  \nonumber  \\
b &=& 0.0025 - 0.0031   \nonumber  \\
c &=& 0.027 - 0.028     \nonumber  \\
d &=& -0.022 - -0.021.
\end{eqnarray}
The values in eq.~(\ref{string}) satisfy the constraints mentioned in eq.~(\ref{abcdnh}).

\subsection{Inverted Hierarchy}
For inverted hierarchy, $\Delta m^2$ is negative and we choose $m_3$ to be
negligibly small leading to $c \approx d$. The ratio of the two
mass-squared differences is
\begin{equation}
\frac{\delta m^2}{|\Delta m^2|} = \frac{6 b (2c+a)}{4c^2 + a^2 + 4b^2} = 0.03.
\end{equation}
This equation is satisfied if
\begin{equation}\label{IHabc}
 a \approx 2c \ {\rm and} \ \frac{b}{c} \approx 0.01.
\end{equation}
The constraint from eq.~(\ref{th12}) forbids the other possibility $b \gg a,c,d$.
From eq.~(\ref{IHabc}), we see that the value of $(c+d-a)$ should
be fine-tuned to $0.5\%$ [$\sim 0.1 (\delta m^2/\Delta m^2$)] to obtain the correct value
of $\theta_{12}$. This is a much more delicate fine-tuning compared to
the NH case.

Demanding that the measured parameters should be within their $3\sigma$
ranges we get the inequalities
\begin{align}\label{IHunpertconstraints}
0.271 \leq \sin^2\theta_{12} \leq 0.345  \nonumber  \\
7.03\times10^{-5} \leq \delta m^2 \leq 8.09\times10^{-5}   \nonumber  \\
-2.635\times10^{-3} \leq \Delta m^2 \leq -2.399\times10^{-3}    \nonumber  \\
|m_3| < 0.1 \ m_1.
\end{align}
This leads to the allowed ranges for $a,b,c$ and $d$
\begin{eqnarray}
a &=& 0.0466 - 0.0506   \nonumber \\
b &=& 0.00024 - 0.00027  \nonumber \\
c &=& 0.0216 - 0.0278    \nonumber \\
d &=& 0.0214 - 0.0278,
\end{eqnarray}
which satisfy the constraints mentioned above. In the case of NH, 
  $b \sim \sqrt{\delta m^2}$ whereas in the case of IH,
$b \sim \delta m^2/a$. Therefore, the value of $b$ in case of IH is
an order of magnitude smaller than in the case of NH, whereas the value
$a$ is an order of magnitude larger than in the case of NH. Note that
the magnitudes of $c$ and $d$ are the same in both cases.

\section{$\mu\leftrightarrow\tau$ symmetry breaking through '$\varepsilon_1$'}

$\mu \leftrightarrow \tau$ symmetry involves two conditions $b_1 = b_2$ and
$c_1 = c_2$, as seen from eq.~(\ref{symmetric}). A violation of either of these conditions leads to a breaking 
of $\mu \leftrightarrow \tau$ symmetry. 
We first consider the breaking of the
condition $b_1 = b_2$. We parametrize this breaking as $b_1 = b - \varepsilon_1$ and 
$b_2 = b + \varepsilon_1$, leading to the neutrino mass matrix,
\begin{equation}\label{pert1}
M_1 = \left(\begin{array}{ccc}
 a               & b - \varepsilon_1  & b + \varepsilon_1  \\
 b - \varepsilon_1  &        c        &       d         \\
 b + \varepsilon_1  &        d        &       c
      \end{array}\right).
\end{equation}
Since $\varepsilon_1$ breaks $\mu\leftrightarrow\tau$ exchange symmetry,
the values of $\theta_{13}$ and $\theta_{23}$ predicted by the 
mass matrix in eq.~(\ref{pert1}) will differ from $0$ and $\pi/4$
respectively.
The characteristic equation for the perturbed mass matrix is 
\begin{equation}
 \lambda^3 - \lambda^2(2c + a) + \lambda(2ca + c^2 - d^2 -2b^2 -2\varepsilon_1^2) - [a(c^2 - d^2) + 2b^2(c + d) + 2\varepsilon_1^2(c - d)] = 0.
\end{equation}
If we impose the condition that the lowest mass eigenvalue is
negligibly small, the quantity in the square brackets in the
above equation should be close to zero. For both NH and IH, we have $c^2 \approx d^2$
 and $b\ll c,d$. Hence the first two terms are negligibly small.
 We require $\varepsilon_1$ to be much less than $c,d$ to satisfy the
 constraint on the lowest mass eigenvalue. 
In this approximation, the characteristic
equation simplifies to
\begin{equation}
 \lambda[\lambda^2 - \lambda(2c + a) + 2ca] = 0,
\end{equation}
whose eigenvalues are $0$, $a$, $2c$. We discuss the cases of
NH and IH separately. 

\subsection{Normal Hierarchy}
For NH, we have $a \approx \sqrt{\delta m^2}$ and $c \approx \sqrt{\Delta m^2}/2$.
The first
element of the eigenvector corresponding to the eigenvalue $m_3$ gives us $\sin\theta_{13}$.
For NH, $m_3 \approx 2c$, and the corresponding eigenvector is
\begin{equation}
 |\nu_3\rangle \approx \left[\begin{array}{c} \frac{\sqrt{2}\varepsilon_1}{(2c-a)}  \\ -\frac{1}{\sqrt{2}} \\ \frac{1}{\sqrt{2}} \end{array} \right].
\end{equation}
The value of $\sin^2\theta_{13}$ can be approximated as $\varepsilon_1^2/(2c^2)$ because $c \gg a$. 
To obtain $\sin^2\theta_{13}\simeq 0.02$, we must have
$\varepsilon_1 \geq b$. Hence we see from eq.~(\ref{pert1}) that
$\varepsilon_1$ can not be treated as a perturbation of the $\mu\leftrightarrow\tau$
symmetric matrix.

We now do a numerical calculation to find the ranges of $a,b,c,d$
and $\varepsilon_1$ allowed by the neutrino oscillation data. 
We find the eigenvalues of matrix in eq.~(\ref{pert1}) and label them
as $m_1, m_2$ and $m_3$ in increasing order. The diagonalizing
matrix is parametrized as
\begin{equation}\label{diagM}
U = \left(\begin{array}{ccc}
U_{e1}    & U_{e2}    & U_{e3} \\
U_{\mu1}  & U_{\mu2}  & U_{\mu3}  \\
U_{\tau1} & U_{\tau2} & U_{\tau3}
       \end{array}\right).
\end{equation}
The 5 oscillation parameters are defined as:
\begin{eqnarray}\label{oscillationparameters}
 \delta m^2 &=& m^2_2 - m^2_1,                             \nonumber  \\
\Delta m^2 &=& m^2_3 - \frac{m^2_1 + m^2_2}{2},           \nonumber  \\
\sin^2\theta_{13} &=& U^2_{e3},                              \nonumber  \\
\sin^2\theta_{23} &=& \frac{U^2_{\mu3}}{1 - U^2_{e3}},       \nonumber  \\
\sin^2\theta_{12} &=& \frac{U^2_{e2}}{1 - U^2_{e3}}.
\end{eqnarray}

\begin{figure}[htbp]
\centering
\begin{minipage}[c]{0.99\textwidth}
\small{(a)}\includegraphics[width=8cm,height=6.5cm,clip]{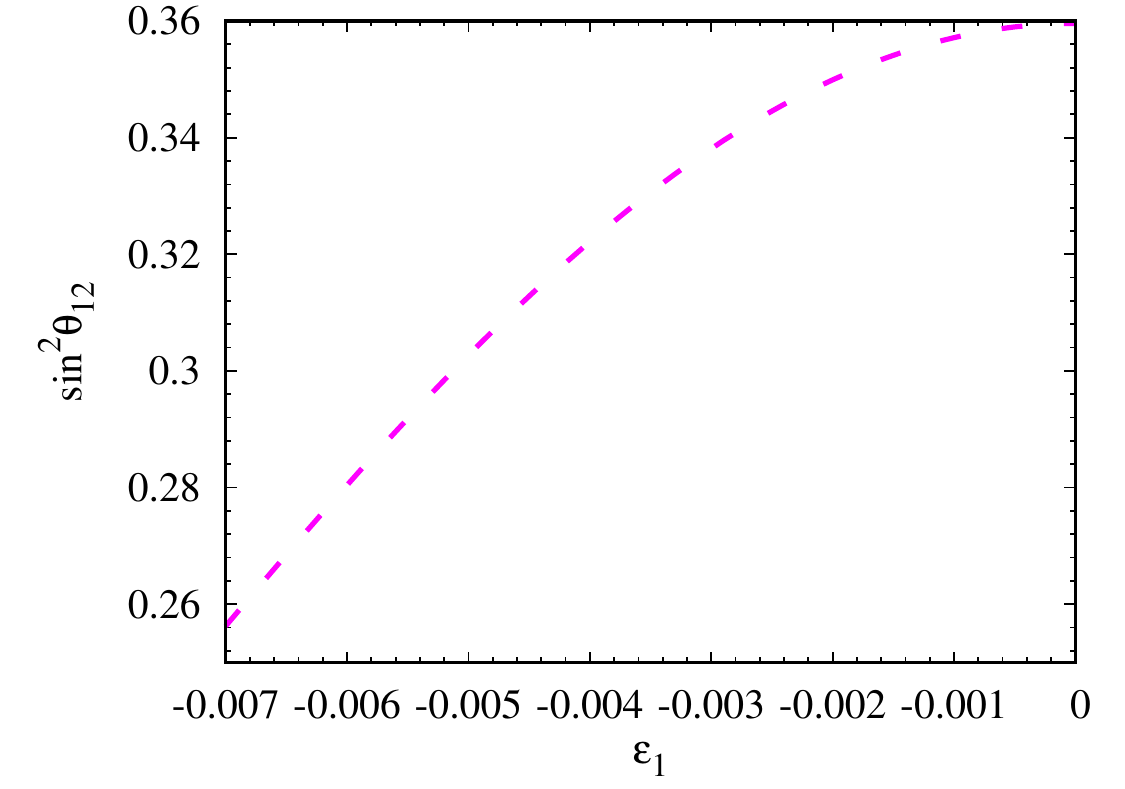}
\hspace{0.1cm}
\small{(b)}\includegraphics[width=8cm,height=6.5cm,clip]{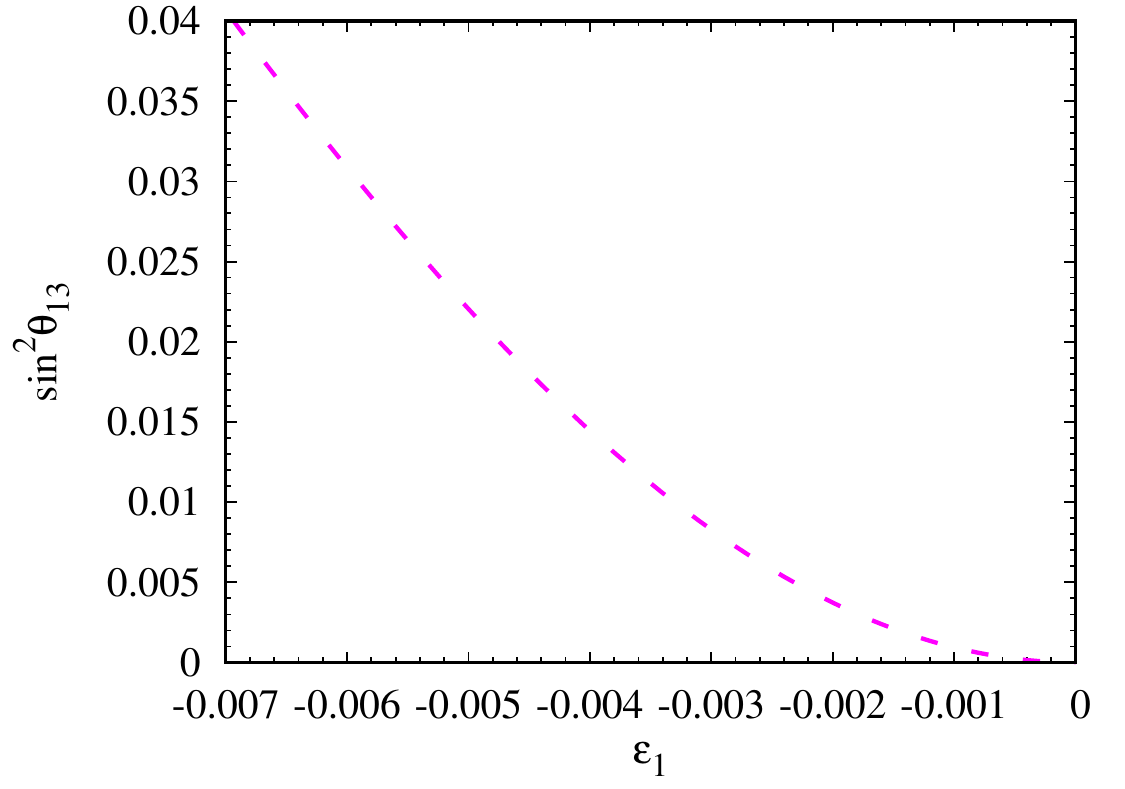}
\small{(c)}\includegraphics[width=8cm,height=6.5cm,clip]{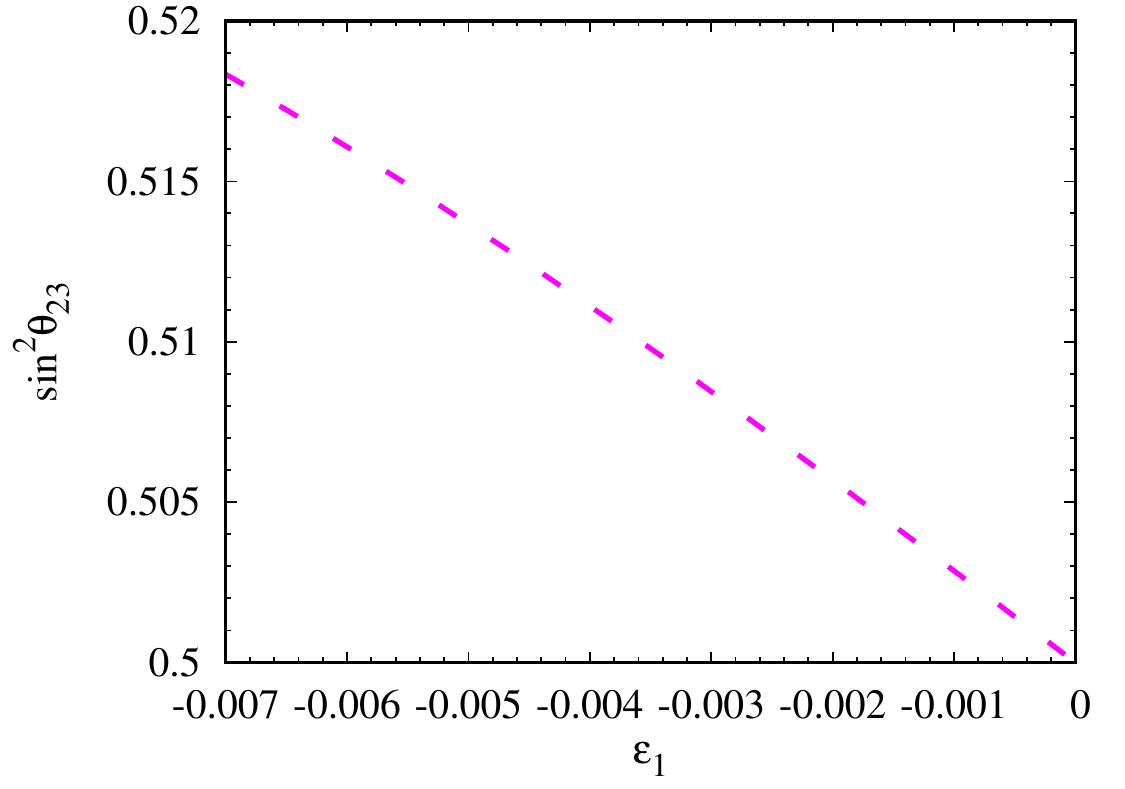}
\end{minipage}
\caption{\label{fig1} Plots of $\sin^2\theta_{ij}$ vs. $\varepsilon_1$ for the central
values of NH neutrino mass matrix elements.}
\end{figure}

As we saw above, the value of $\varepsilon_1$ needed to generate the
correct magnitude of $\theta_{13}$ means $\varepsilon_1 \geq b$.
Therefore, we treat
$\varepsilon_1$ as a free parameter and numerically search for allowed values
of $a, b, c, d$ and $\varepsilon_1$ which satisfy the following $3\sigma$
experimental constraints on $\sin^2\theta_{13}$ and $\sin^2\theta_{23}$
\begin{align}\label{pert1constraints}
0.385 \leq \sin^2\theta_{23} \leq 0.635     \nonumber  \\
 0.01934 \leq \sin^2\theta_{13} \leq 0.02392,
\end{align}
in addition to the four constraints already given in eq.~(\ref{NHunpertconstraints}).
Our numerical search gives
\begin{eqnarray}
a &=& 0.0027 - 0.0046   \nonumber \\
b &=& 0.0026 - 0.0032  \nonumber \\
c &=& 0.028             \nonumber \\
d &=& -0.022        \nonumber  \\
\varepsilon_1 &=& -0.0053 - -0.0046.
\end{eqnarray}
For the central value $\varepsilon_1 = -0.0050$, we get $\sin^2\theta_{12}=0.298$,
$\sin^2\theta_{13} = 0.0221$
and $\sin^2\theta_{23} = 0.514$. The T2K experiment observes maximal 
$\nu_\mu$ disappearance, implying $|U_{\mu 3}|^2=0.5=\cos^2\theta_{13}\sin^2\theta_{23}$.
On substituting the reactor measurements of $\theta_{13}$, this leads to
$\sin^2\theta_{23}=0.514$, which is equal to the prediction above.
It is interesting to note that the value of $\varepsilon_1$, needed to
produce the correct value of $\sin^2\theta_{13}$ also produces the
correct deviation in $\sin^2\theta_{23}$ needed to explain the T2K
$\nu_\mu$ disappearance data. The variation of $sin^2\theta_{ij}$ vs. $\varepsilon_1$
is plotted in fig.~\ref{fig1}, for NH. It was mentioned earlier that a fine-tuning
of neutrino mass matrix parameters is required to obtain viable values of $\theta_{12}$.
There is a significant variation of $\sin^2\theta_{12}$ with respect to $\varepsilon_1$
because of this fine-tuning. As shown above, $\sin^2\theta_{13}$ varies as $\varepsilon_1^2$.
We see that $\sin^2\theta_{23}$ shows a small linear variation with respect to $\varepsilon_1$.

\subsection{Inverted Hierarchy}
For IH, $m_3 \approx 0$, whose eigenvector is 
\begin{equation}
 |\nu_3\rangle \approx \left[\begin{array}{c} -\frac{\sqrt{2}\varepsilon_1}{a}  \\ -\frac{1}{\sqrt{2}} \\ \frac{1}{\sqrt{2}} \end{array} \right].
\end{equation}
Hence $\sin^2\theta_{13} = 2\varepsilon_1^2/a^2$. Since the
value of $a$ in IH is the same as the value of $2c$ in NH ($\approx\sqrt{\Delta m^2}$),
the magnitude of $\varepsilon_1$ in this case is of a similar
magnitude as that of NH. But the value of $b$ in IH is an order of
magnitude lower than the case of NH and hence we have $b\ll\varepsilon_1$
in the case of IH. Here $\varepsilon_1$ most definitely can not be
treated as a perturbation on $b$.

\begin{figure}[htbp]
\begin{minipage}[c]{0.99\textwidth}
\small{(a)}\includegraphics[width=8cm,height=6.5cm,clip]{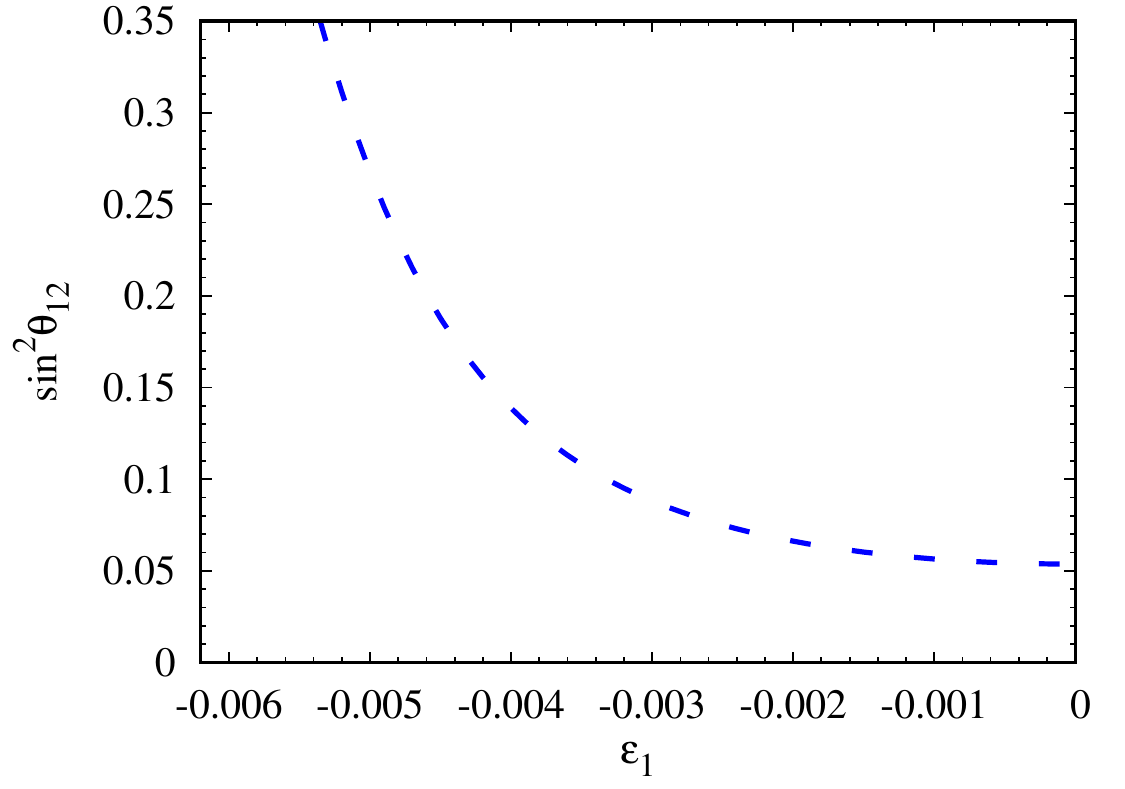}
\hspace{0.1cm}
\small{(b)}\includegraphics[width=8cm,height=6.5cm,clip]{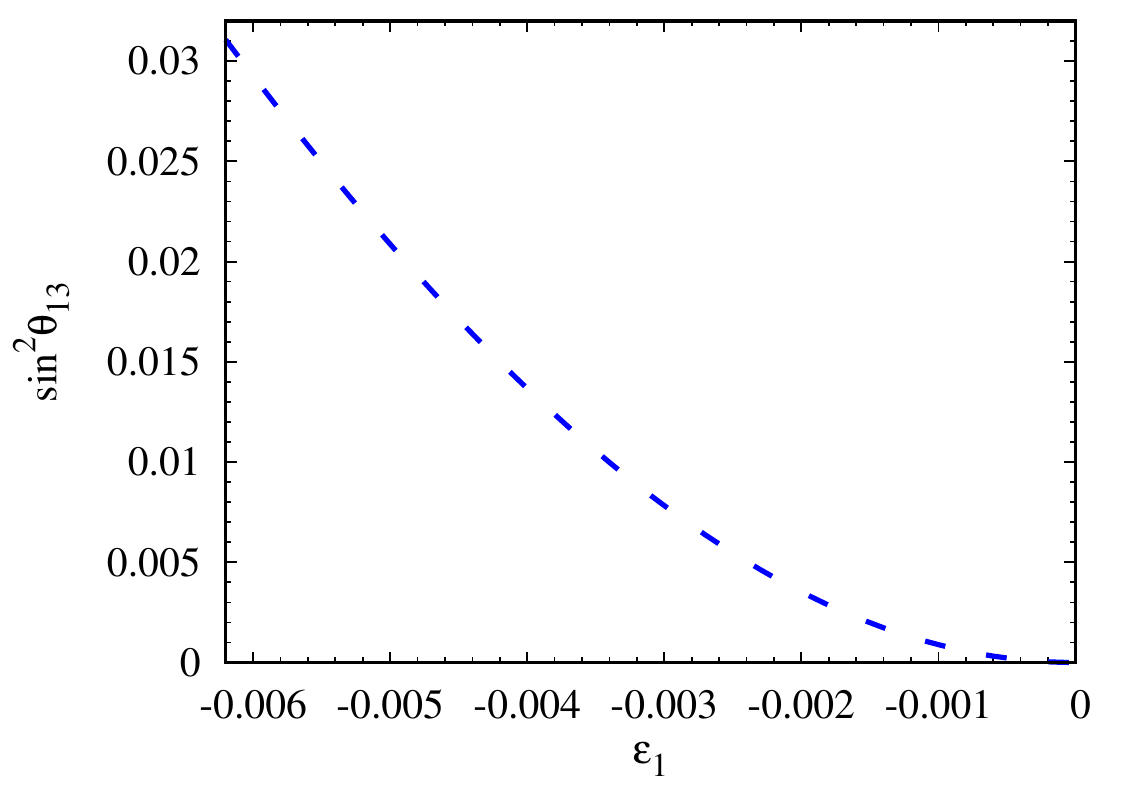}
\small{(c)}\includegraphics[width=8cm,height=6.5cm,clip]{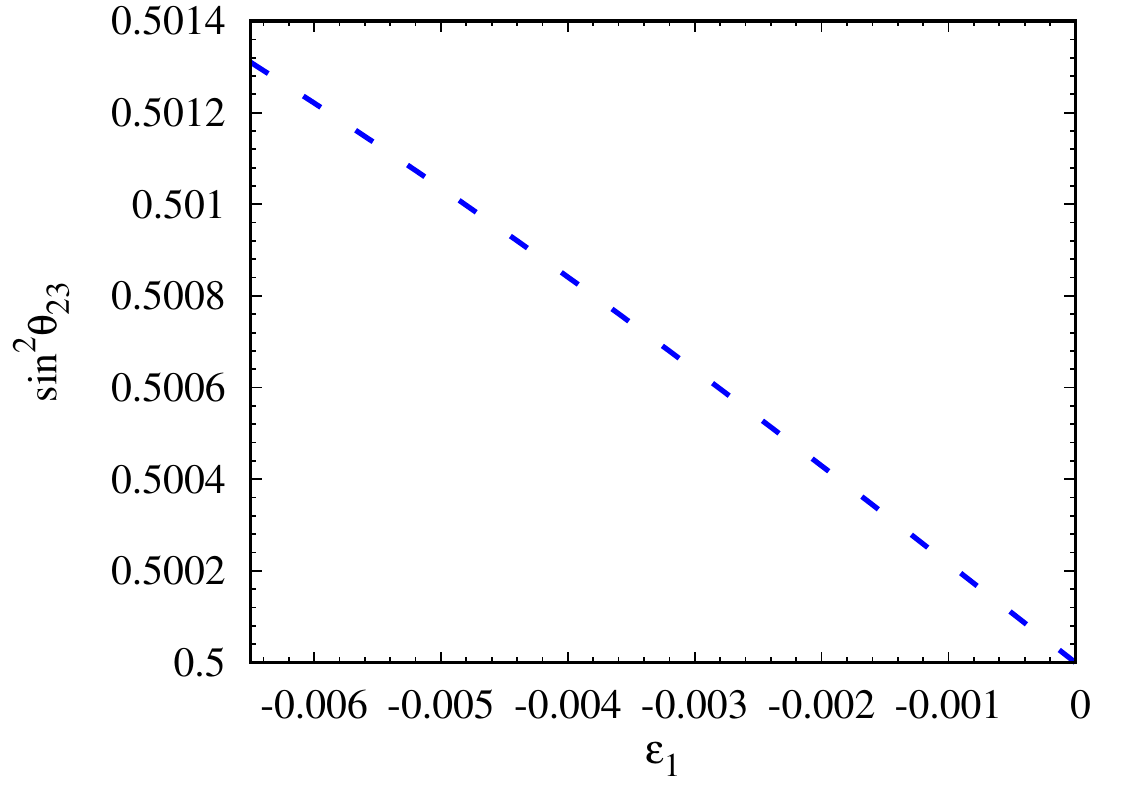}
\end{minipage}
\caption{\label{fig2} Plots of $\sin^2\theta_{ij}$ vs. $\varepsilon_1$ for the central
values of IH neutrino mass matrix elements.}
\end{figure}

For IH, the lowest eigenvalue of the matrix in eq.~(\ref{pert1})
is labeled $m_3$, the middle one is labeled $m_1$ and the highest
$m_2$. The diagonalizing matrix is labeled as in eq.~(\ref{diagM})
and the definitions of the five oscillation parameters remain
the same as those in eq.~(\ref{oscillationparameters}). For this
case also we do a numerical search to find ranges of $a, b, c, d$
and $\varepsilon_1$ which satisfy the six experimental constraints
given in eqs.~(\ref{IHunpertconstraints}) and~(\ref{pert1constraints}). The search yields the
ranges
\begin{eqnarray}
a &=& 0.048 - 0.050   \nonumber \\
b &=& 0.00022 - 0.00027  \nonumber \\
c &=& 0.023 - 0.026     \nonumber \\
d &=& 0.0243 - 0.0277        \nonumber  \\
\varepsilon_1 &=& -0.0061 - -0.0049.
\end{eqnarray}
For the central value $\varepsilon_1 = -0.0052$, we get $\sin^2\theta_{12}=0.306$,
$\sin^2\theta_{13} = 0.0223$
and $\sin^2\theta_{23} = 0.501$.
The variation of $sin^2\theta_{ij}$ vs. $\varepsilon_1$
is plotted in fig.~\ref{fig2}, for IH.
Since $b$ is too small, extreme fine-tuning is needed to obtain
the appropriate value of $\sin^2\theta_{12}$. The variation of $\sin^2\theta_{12}$,
with respect to $\varepsilon_1$, is very pronounced because of this extreme
fine-tuning. As in the case of NH, $\sin^2\theta_{13}$ varies as $\varepsilon_1^2$
and $\sin^2\theta_{23}$ shows a small linear variation with respect to $\varepsilon_1$.

\section{$\mu\leftrightarrow\tau$ symmetry breaking through '$\varepsilon_2$'}

Now we hold the equality $b_1=b_2=b$ in eq.~(\ref{symmetric}) but assume
$c_1 \neq c_2$. We parametrize this breaking of $\mu \leftrightarrow \tau$ 
symmetry as $c_1=c-\varepsilon_2$ and $c_2=c+\varepsilon_2$. The neutrino
mass matrix has the form
\begin{equation}\label{pert2}
M_2 = \left(\begin{array}{ccc}
 a & b              & b               \\
 b & c - \varepsilon_2 & d               \\
 b & d              & c + \varepsilon_2
      \end{array}\right).
\end{equation}
The $2-3$ block is diagonalized by
applying the similarity transformation $U_{23}^TM_2U_{23}$, where
\begin{equation}\label{U23}
U_{23}=\left( \begin{array}{ccc}
1 & 0 & 0  \\
0 & \cos\theta_{23} &  \sin\theta_{23}  \\
0 & -\sin\theta_{23} &  \cos\theta_{23}
\end{array} \right).
\end{equation}
In the above equation $\theta_{23}\neq -\pi/4$ but is taken to be $-\pi/4+\delta\theta_{23}$.
The deviation from maximality is found to be
\begin{equation}
\delta\theta_{23}  \simeq -\frac{\varepsilon_2}{2d}.
\end{equation}
From neutrino data in Table~1, 
the maximum allowed value of this quantity is $0.12$ \cite{Esteban:2016qun}. 
The $13$ element of the rotated mass matrix is $\sqrt{2}b\varepsilon_2/(2d)$.
This term determines the value of $\sin\theta_{13}$. (The corresponding
quantity in the case of $\varepsilon_1$ symmetry breaking is $\sqrt{2}\varepsilon_1$).
Given the limit on $\delta\theta_{23}$, we find 
$b\varepsilon_2/(\sqrt{2}d)$ is an order of magnitude smaller than $b$.

In the earlier discussion on $\varepsilon_1$ symmetry breaking, it was shown
that $\varepsilon_1 \simeq 0.005$ to reproduce the correct $\sin^2\theta_{13}$.
Therefore the term generating non-zero $\theta_{13}$ for $\varepsilon_2$
symmetry breaking is an order of magnitude
lower for NH ($b\sim 0.002$) and two orders of magnitude lower
for IH ($b\sim 0.0002$).
Hence it is impossible to satisfy the constraints on $\sin^2\theta_{13}$
and $\sin^2\theta_{23}$ only through $\varepsilon_2$ breaking.
The maximum allowed values of $\sin^2\theta_{13}$ we get in this case
are of the order of $10^{-5}$ for NH and $10^{-7}$ for IH.

\section{Complete $\mu\leftrightarrow\tau$ Symmetry Breaking}

In the above sections, we saw that $\varepsilon_1$ symmetry
breaking generates acceptable value of $\sin^2\theta_{13}$ but
keeps the value of $\sin^2\theta_{23}$ close to maximal.
On the other hand, $\varepsilon_2$ symmetry breaking leads to
significant deviation of $\sin^2\theta_{23}$ away from
maximality but predicts very small values for $\sin^2\theta_{13}$.
If later data         
confirms that $\sin^2\theta_{23}$ is non-maximal, we need to
introduce both $\varepsilon_1$ and $\varepsilon_2$ symmetry breaking
  to describe the neutrino mixing angles accurately. In such a
situation, the value of $\varepsilon_1$ is essentially determined
by $\sin^2\theta_{13}$ and that of $\varepsilon_2$ is determined
by the deviation of $\theta_{23}$ from maximality.

The neutrino mass matrix is described by six parameters: $a,b,c,d,
\varepsilon_1$ and $\varepsilon_2$. We search for the allowed values
of these parameters by demanding that the two mass-squared
differences and the three mixing angles should be within their
allowed $3\sigma$ ranges. We also impose a sixth constraint that
lowest neutrino mass ($m_1$ for NH and $m_3$ for IH) 
should be less than $0.001$ eV. With these constraints we obtain
the following allowed ranges of parameters.
For NH
\begin{eqnarray}\label{completeNH}
 a&=&0.0027-0.0046 \\ \nonumber
 |b|&=&0.0026-0.0038 \\ \nonumber
c&=&0.028 \\ \nonumber
d&=&-0.022       \\ \nonumber
|\varepsilon_1|&=&0.0043-0.0052\\ \nonumber
|\varepsilon_2|&=&0.0-0.0046.
\end{eqnarray}
Similarly for IH,
\begin{eqnarray}\label{completeIH}
 a&=&0.0048-0.0050 \\ \nonumber
|b|&=&0.0-0.00027 \\ \nonumber
c&=&0.023-0.028 \\ \nonumber
d&=&0.0210-0.0270  \\ \nonumber
|\varepsilon_1|&=&0.0044-0.0058    \\   \nonumber
 |\varepsilon_2|&=&0.0-0.0026.
\end{eqnarray}
Earlier we saw the following patterns: for NH $d\approx-c$,
$a\approx b\approx \varepsilon_1$ and for IH $c\approx d\approx a/2$,
$b\ll\varepsilon_1$ with $\varepsilon_1\rm{(NH)}=\varepsilon_1\rm{(IH)}$.
From above equations we see that the same relations hold here also.

\section{Ranges of neutrino mass matrix parameters from precision oscillation data}
In the previous sections we varied the parameters of neutrino mass matrix
to find their values which satisfy the experimental constraints.
As we can see from eqs.~(\ref{completeNH}) and (\ref{completeIH}),
the ranges for these parameters are quite small. In this section,
we do a systematic search to find the exact ranges of these parameters
allowed by the current oscillation data. Among the oscillation observables,
the mass-squared differences, 
$\delta m^2$ \cite{Gando:2010aa}and $|\Delta m^2|$ \cite{Adamson:2013whj, Abe:2011sj, Adamson:2017gxd}, 
  are measured to better than $3\%$ precision. The mixing angles,
$\sin^2\theta_{12}$ \cite{Abdurashitov:1994bc, Hampel:1998xg, Abe:2016nxk, Ahmad:2001an, Bahcall:2001zu} and    
 $\sin^2\theta_{13}$ \cite{An:2012eh}, are determined to about $4\%$                    
precision. The precision in $\sin^2\theta_{23}$ is poorer because
of the octant ambiguity \cite{Adamson:2013whj, Abe:2011sj, Adamson:2017gxd}. Below we study         
the impact of these precision measurements on the allowed ranges of neutrino mass
parameters.

We use the following procedure. 
We first choose a value for the lowest neutrino
mass eigenvalue.
We then choose five uniform random numbers in the
interval [-1,1]. Using these numbers, we construct random
values for the five neutrino oscillation parameters within
their $1\ \sigma$ ranges. We construct the diagonal neutrino
mass matrix using the lowest neutrino mass and the two mass-squared
differences. For NH, the diagonal form of the mass matrix is
\begin{equation}
M_{\rm diag}=\left( \begin{array}{ccc}
m_1      &                           &                                           \\
         & \sqrt{m_1^2 + \delta m^2} &                                           \\
         &                           & \sqrt{m_1^2 + \delta m^2/2 + \Delta m^2}  \end{array} \right),
\end{equation}
where $m_1$ is the lowest neutrino mass chosen, whereas for IH, this
matrix takes the form
\begin{equation}
M_{\rm diag}=\left( \begin{array}{ccc}
\sqrt{m_3^2 - \delta m^2/2 + |\Delta m^2|}   &                                            &           \\
                                           & \sqrt{m_3^2 + \delta m^2/2 + |\Delta m^2|}   &           \\
                                           &                                            & m_3  \end{array} \right),
\end{equation}
where $m_3$ is the lowest neutrino mass. We obtain the neutrino
mass matrix in flavor basis by the similarity transformation 
$M_0=UM_{\rm diag}U^T$, where $U$ is the orthogonal matrix constructed
using the values of the three chosen mixing angles.

For a given set of five random numbers we get the corresponding
set of neutrino oscillation parameters which in turn lead to a given
set of values for $a,b,c,d,\varepsilon_1$ and $\varepsilon_2$. We repeat
this procedure for 10,000 sets of five random numbers to produce
10,000 values of neutrino mass matrix parameters. From these 10,000
sets of parameter values we tabulate the mean, the standard deviation,
the lowest and the highest values. This procedure is used to construct the
allowed ranges of neutrino mass matrix elements
for the following
eight cases: for NH, $m_1=0,0.001,0.01$ and $0.1$ eV and for IH,
$m_3=0,0.001,0.01$ and $0.1$ eV.

From these tables we note the following patterns. The ranges for the neutrino mass matrix
elements, whose magnitudes are large, are  $\lesssim1\%$. This is true
for the parameters $c$ and $d$ in all cases and for the parameter $a$ in the case
of IH and when the minimum neutrino mass $m_1\geq 0.01$ eV in the case of NH.
Since $\sin^2\theta_{13}\propto\varepsilon_1^2$, the range of $\varepsilon_1$ is about
$2\%$, which is half the uncertainty in $\sin^2\theta_{13}$. The range of $\varepsilon_2$
is about $10\%$ in case of NH and about $25\%$ in case of IH. The values and ranges of $b$
are usually very small because of the need to otain the correct value
of $\theta_{12}$. 

\begin{table}[htbp]
\centering
\label{NH0}
\begin{tabular}{|c|c|c|c|c|}
\hline
Matrix          & Lower          & Upper         & Mean       & Standard     \\
Element         & Bound          & Bound         &            & Deviation    \\ \hline
$a$             & 0.005426       & 0.006546      & 0.005971   & 0.0002192    \\ \hline 
$b$             & -0.001934      & -0.001568     & -0.001752  & 0.00007519   \\ \hline
$c$             & 0.02621        & 0.02708       & 0.02665    & 0.0001534    \\ \hline
$d$             & -0.02007       & -0.01941      & -0.01974   & 0.0001215    \\ \hline
$\varepsilon_1$    & 0.008661       & 0.009855      & 0.009252   & 0.0002200    \\ \hline
$\varepsilon_2$    & 0.003840       & 0.005704      & 0.004773   & 0.0004379    \\ \hline
\end{tabular}
\caption{Normal Hierarchy: $m_1=0.0$ eV}
\end{table}

\begin{table}[htbp]
\centering
\label{NH-3}
\begin{tabular}{|c|c|c|c|c|}
\hline
Matrix          & Lower          & Upper         & Mean       & Standard     \\
Element         & Bound          & Bound         &            & Deviation    \\ \hline
$a$             & 0.006123       & 0.007233      & 0.006665   & 0.0002130    \\ \hline 
$b$             & -0.001684      & -0.001298     & -0.001484  & 0.00007498   \\ \hline
$c$             & 0.02639        & 0.02729       & 0.02684    & 0.0001495    \\ \hline
$d$             & -0.01996       & -0.01932      & -0.01964   & 0.0001208    \\ \hline
$\varepsilon_1$    & 0.008506       & 0.009695      & 0.009089   & 0.0002174    \\ \hline
$\varepsilon_2$    & 0.003708       & 0.005507      & 0.004630   & 0.0004337    \\ \hline
\end{tabular}
\caption{Normal Hierarchy: $m_1=0.001$ eV}
\end{table}

\begin{table}[htbp]
\centering
\label{NH-2}
\begin{tabular}{|c|c|c|c|c|}
\hline
Matrix          & Lower          & Upper         & Mean       & Standard     \\
Element         & Bound          & Bound         &            & Deviation    \\ \hline
$a$             & 0.01352        & 0.01443       & 0.01395    & 0.0001761    \\ \hline 
$b$             & -0.0004328     & -0.00008957   & -0.0002615 & 0.00006912   \\ \hline
$c$             & 0.03005        & 0.03082       & 0.03043    & 0.0001347    \\ \hline
$d$             & -0.01797       & -0.01738      & -0.01768   & 0.0001167    \\ \hline
$\varepsilon_1$    & 0.007172       & 0.008227      & 0.007688   & 0.0001947    \\ \hline
$\varepsilon_2$    & 0.002904       & 0.004506      & 0.003707   & 0.0003989    \\ \hline
\end{tabular}
\caption{Normal Hierarchy: $m_1=0.01$ eV}
\end{table}

\begin{table}[htbp]
\centering
\label{NH-1}
\begin{tabular}{|c|c|c|c|c|}
\hline
Matrix          & Lower          & Upper         & Mean       & Standard     \\
Element         & Bound          & Bound         &            & Deviation    \\ \hline
$a$             & 0.1009         & 0.1012        & 0.1010     & 0.00005170   \\ \hline 
$b$             & 0.00004479     & 0.0001452     & 0.00009461 & 0.00002089   \\ \hline
$c$             & 0.1056         & 0.1059        & 0.1057     & 0.00005338   \\ \hline
$d$             & -0.005459      & -0.005226     & -0.005343  & 0.00004978   \\ \hline
$\varepsilon_1$    & 0.002074       & 0.002424      & 0.002245   & 0.00006093   \\ \hline
$\varepsilon_2$    & 0.0007954      & 0.001298      & 0.001044   & 0.0001206    \\ \hline
\end{tabular}
\caption{Normal Hierarchy: $m_1=0.1$ eV}
\end{table}

\begin{table}[htbp]
\centering
\label{IH0}
\begin{tabular}{|c|c|c|c|c|}
\hline
Matrix          & Lower                 & Upper         & Mean       & Standard     \\
Element         & Bound                 & Bound         &            & Deviation    \\ \hline
$a$             & 0.04318               & 0.04504       & 0.04413    & 0.0003216    \\ \hline 
$b$             & $4.258\times10^{-6}$  & 0.0005179     & 0.0002587  & 0.0001321    \\ \hline
$c$             & 0.02810               & 0.02909       & 0.02859    & 0.0001770    \\ \hline
$d$             & 0.02114               & 0.02218       & 0.02168    & 0.0001833    \\ \hline
$\varepsilon_1$    & -0.01176              & -0.01062      & -0.01120   & 0.0002125    \\ \hline
$\varepsilon_2$    & 0.0009376             & 0.002888      & 0.001915   & 0.0004803    \\ \hline
\end{tabular}
\caption{Inverted Hierarchy: $m_3=0.0$ eV }
\end{table}

\begin{table}[htbp]
\centering
\label{IH-3}
\begin{tabular}{|c|c|c|c|c|}
\hline
Matrix          & Lower                  & Upper         & Mean       & Standard     \\
Element         & Bound                  & Bound         &            & Deviation    \\ \hline
$a$             & 0.04319                & 0.04501       & 0.04412    & 0.0003206    \\ \hline 
$b$             & $-6.945\times10^{-7}$  & 0.0005158     & 0.0002593  & 0.0001309    \\ \hline
$c$             & 0.02809                & 0.02912       & 0.02859    & 0.0001793    \\ \hline
$d$             & 0.02115                & 0.02217       & 0.02167    & 0.0001826    \\ \hline
$\varepsilon_1$    & -0.01179               & -0.01061      & -0.01120   & 0.0002152    \\ \hline
$\varepsilon_2$    & 0.0009429              & 0.002899      & 0.001917   & 0.0004754    \\ \hline
\end{tabular}
\caption{Inverted Hierarchy: $m_3=0.001$ eV}
\end{table}

\begin{table}[htbp]
\centering
\label{IH-2}
\begin{tabular}{|c|c|c|c|c|}
\hline
Matrix          & Lower          & Upper         & Mean       & Standard     \\
Element         & Bound          & Bound         &            & Deviation    \\ \hline
$a$             & 0.04526        & 0.04685       & 0.04606    & 0.0002858    \\ \hline 
$b$             & -0.00003361    & 0.0004061     & 0.0001856  & 0.0001106    \\ \hline
$c$             & 0.03264        & 0.03355       & 0.03310    & 0.0001632    \\ \hline
$d$             & 0.01770        & 0.01859       & 0.01815    & 0.0001604    \\ \hline
$\varepsilon_1$    & -0.009867      & -0.008859     & -0.009364  & 0.0001802    \\ \hline
$\varepsilon_2$    & 0.0007994      & 0.002452      & 0.001618   & 0.0004015    \\ \hline
\end{tabular}
\caption{Inverted Hierarchy: $m_3=0.01$ eV}
\end{table}

\begin{table}[htbp]
\centering
\label{IH-1}
\begin{tabular}{|c|c|c|c|c|}
\hline
Matrix          & Lower          & Upper         & Mean       & Standard     \\
Element         & Bound          & Bound         &            & Deviation    \\ \hline
$a$             & 0.1101         & 0.1107        & 0.1104     & 0.0001094    \\ \hline 
$b$             & -0.00004422    & 0.00008396    & 0.00001912 & 0.00003190   \\ \hline
$c$             & 0.1065         & 0.1068        & 0.1067     & 0.00006381   \\ \hline
$d$             & 0.005095       & 0.005409      & 0.005255   & 0.00005902   \\ \hline
$\varepsilon_1$    & -0.002854      & -0.002540     & -0.002696  & 0.00005532   \\ \hline
$\varepsilon_2$    & 0.0002519      & 0.0007285     & 0.0004873  & 0.0001155    \\ \hline
\end{tabular}
\caption{Inverted Hierarchy: $m_3=0.1$ eV}
\end{table}

\section{$\mu\leftrightarrow-\tau$ symmetry}
From the tables given in the previous section, we note that $b=0$
is an accepted value for the case of IH. In this section, we explore
the allowed values of neutrino mass matrix with the constraint $b\equiv0$.
It is possible to impose such a constraint through $\mu\leftrightarrow-\tau$
exchange symmetry. Under this symmetry, the $\varepsilon_1$ term is naturally
non-zero.

The most general neutrino mass matrix invariant under this symmetry is
\begin{equation}\label{IHb=0}
M_3 = \left(\begin{array}{ccc}
 a               & -\varepsilon_1     & \varepsilon_1      \\
 -\varepsilon_1     &        c        &       d         \\
 \varepsilon_1      &        d        &       c
\end{array}\right).
\end{equation}
Diagonalizing this matrix, we find $\theta_{23}=-\pi/4$, $\theta_{12}=0$
and
\begin{equation}\label{tan2t13prime}
 \tan2\theta_{13}=\frac{2\sqrt{2}\varepsilon_1}{c-d-a}\approx-\frac{2\sqrt{2}\varepsilon_1}{a},
\end{equation}
because $c\approx d$ for IH. Also, we note that $a\approx 2c$.
This, except for $\theta_{23}$, is exactly opposite to $\mu\leftrightarrow\tau$
symmetry case where we had $\theta_{13}=0$  and $\tan2\theta_{12}=2\sqrt{2}b/(c+d-a)$.
Since $\theta_{13} \ll 1$, the above equation implies that $\varepsilon_1\ll a,c,d$.
Obviously, $\mu\leftrightarrow-\tau$ is not exact because it predicts
$\theta_{12}=0$. It can be broken through $\varepsilon_2$ term introduced in $22$ and
$33$ elements as in the case of $\mu\leftrightarrow\tau$ symmetry. We will show below
that such a breaking can lead to both non-maximal $\theta_{23}$ as well as
viable values of $\theta_{12}$. However to obtain $\theta_{12}$ within 
the experimentally allowed range, we need to fine-tune the combination
$c+d-a$ to order $\varepsilon_1^2/a$.

\begin{figure}[htbp]
\begin{minipage}[c]{0.99\textwidth}
\small{(a)}\includegraphics[width=8cm,height=6.5cm,clip]{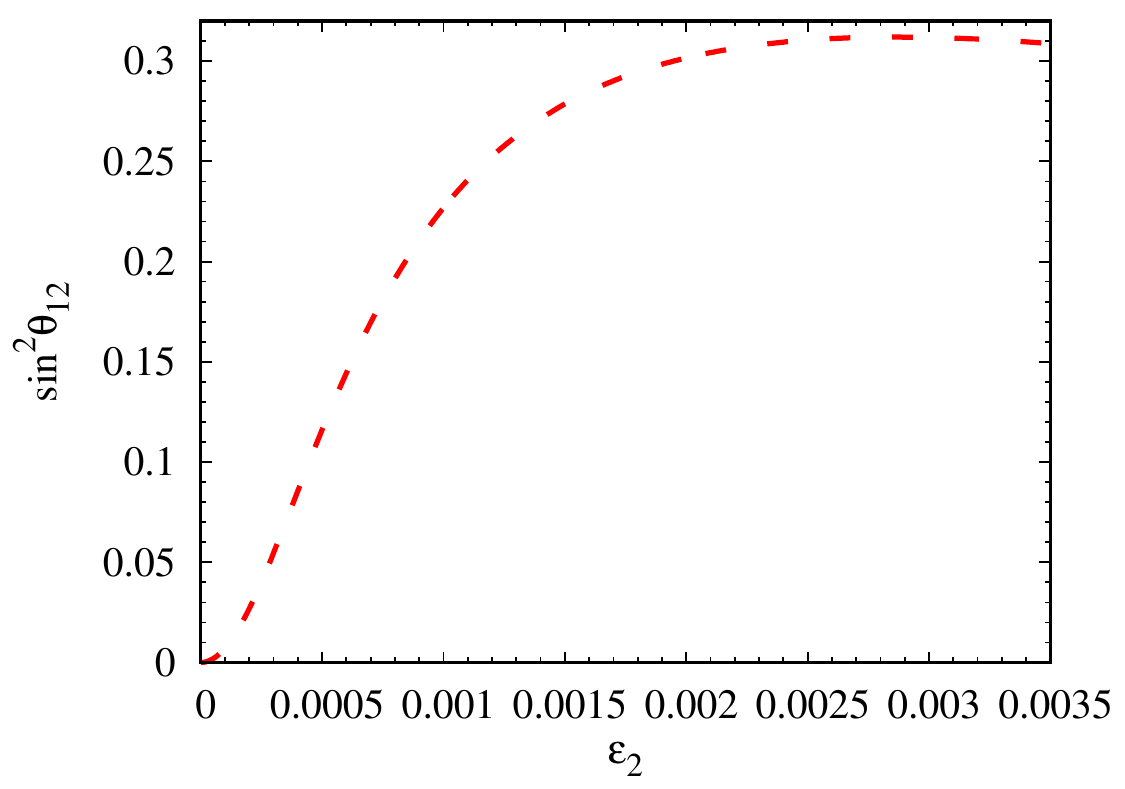}
\hspace{0.1cm}
\small{(b)}\includegraphics[width=8cm,height=6.5cm,clip]{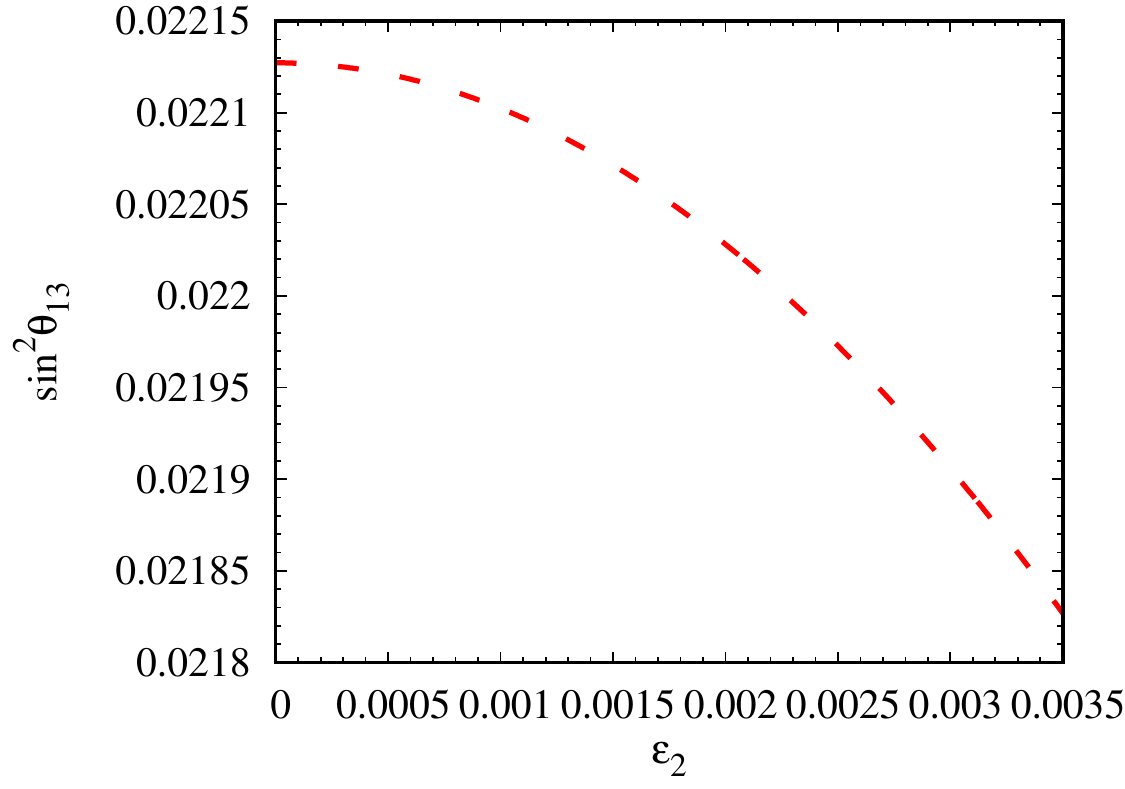}
\end{minipage}
\begin{center}
\small{(c)}\includegraphics[width=8cm,height=6.5cm,clip]{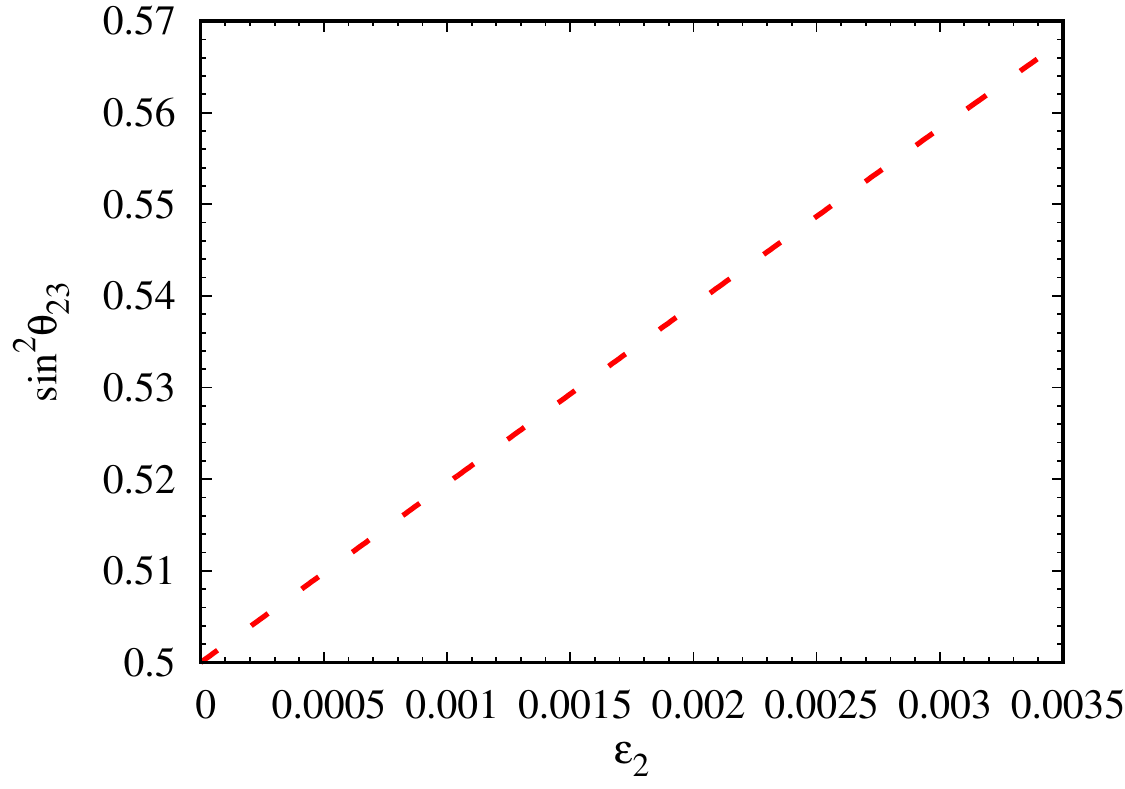} 
\end{center}
\caption{\label{fig3}Plots of $\sin^2\theta_{ij}$ vs. $\varepsilon_2$ for the central
values of IH ($b\equiv0$) neutrino mass matrix elements.}
\end{figure}

With the $\varepsilon_2$ symmetry breaking the neutrino mass matrix becomes
\begin{equation}\label{IHb=0pert}
M_4 = \left(\begin{array}{ccc}
 a               & -\varepsilon_1           & \varepsilon_1      \\
 -\varepsilon_1     &        c-\varepsilon_2   &       d         \\
     \varepsilon_1  &        d              &  c+\varepsilon_2
      \end{array}\right).
\end{equation}
Applying the similarity transformation $U_{23}^TM_4U_{23}$, where
$U_{23}$ is defined in eq.~(\ref{U23}), we get
\begin{equation}
U^T_{23}M_4U_{23}=\left( \begin{array}{ccc}
a                                           &   -\sqrt{2}\varepsilon_1\sin\delta\theta_{23}     &  \sqrt{2}\varepsilon_1\cos\delta\theta_{23}         \\
-\sqrt{2}\varepsilon_1\sin\delta\theta_{23}    &   c+d\sqrt{1+\frac{\varepsilon_2^2}{d^2}}         &  0                                         \\
\sqrt{2}\varepsilon_1\cos\delta\theta_{23}      &    0                                           &  c-d\sqrt{1+\frac{\varepsilon_2^2}{d^2}} \end{array} \right).
\end{equation}
Here, $\delta\theta_{23}$ is the deviation of $\theta_{23}$ from maximality
and it is given by $\tan2\delta\theta_{23}=-\varepsilon_2/d$. Note that
the $12$ element of this matrix is proportional to $\varepsilon_1\varepsilon_2$.
We now apply a further similarity transformation through the orthogonal
matrix
\begin{equation}
U_{13}U_{12} = \left( \begin{array}{ccc}
c_{13}c_{12}   &    c_{13}s_{12}       &   s_{13}      \\
-s_{12}        &    c_{12}             &   0           \\
-s_{13}c_{12}  &    -s_{13}s_{12}      &   c_{13}     \end{array} \right).
\end{equation}
We demand that the $13$ and $23$ elements of the transformed matrix to be zero.
The explicit expressions for these elements are given in Appendix.
If we neglect terms which are third order in the small quantities $\varepsilon_1$
and $\varepsilon_2$, both these conditions lead to
\begin{equation}
  \tan2\theta_{13}=\frac{2\sqrt{2}\varepsilon_1\cos\delta\theta_{23}}{c-d'-a}\approx-\frac{2\sqrt{2}\varepsilon_1}{a},
\end{equation}
where $d'=d\sqrt{1+\varepsilon_2^2/d^2}$ and we set $\cos\delta\theta_{23}\approx 1$.
This is very similar to the relation we had for the exact
$\mu\leftrightarrow-\tau$ symmetry case, as given in eq.~(\ref{tan2t13prime}).
Note that the value of $\varepsilon_2$ is fixed by the measured value
of $\delta\theta_{23}$ and that of $\varepsilon_1$ by $\theta_{13}$.
Viable values of $\theta_{12}$ can be obtained by fine-tuning the
combination $c+d'-a$. 
Demanding the $12$ element of the transformed matrix to be zero,
we get
\begin{equation}
 \tan2\theta_{12}\approx-\frac{a}{d}\frac{\sqrt{2}\varepsilon_1\varepsilon_2}{a(a-c-d')+4\varepsilon_1^2}.
\end{equation}
By fine-tuning $(a-c-d')\sim \varepsilon_1^2/a$, it is possible
to obtain $\sin^2\theta_{12}\approx 0.3$. The variation of $\sin^2\theta_{ij}$
with respect to $\varepsilon_2$ is plotted in fig.~\ref{fig3}. As in the case of
$\mu\leftrightarrow\tau$ symmetry for IH, there is little variation of $\sin^2\theta_{13}$
and a linear variation of $\sin^2\theta_{23}$. The variation of $\sin^2\theta_{12}$
is quite sharp because of the fine-tuning of $(a-c-d')$.
This fine-tuning
does not have a significant effect on the neutrino mass eigenvalues
which determine the values of $a,c,d$ originally. Thus it is possible
to predict all the neutrino oscillation parameters with a single
breaking of $\mu\leftrightarrow-\tau$ symmetry through $\varepsilon_2$.

\section{Conclusion}\label{Conclusion}
We have considered the constraints imposed by the precision oscillation
data on $\mu\leftrightarrow\tau$ symmetric neutrino mass matrix. We find that
the elements of this matrix are confined to be in extremely narrow ranges by
the current data, both for normal hierarchy and for inverted hierarchy. 
There are two parameters which break the $\mu\leftrightarrow\tau$ symmetry,
$\varepsilon_1$ and $\varepsilon_2$. Even though $\varepsilon_1$ is small, it can not be
treated as a perturbation because its value is comparable (for NH) or
much larger than (for IH) the relevant element of neutrino mass matrix.
A value of $\varepsilon_1\sim0.005$ eV (for both NH and IH) 
leads to a viable value of $\theta_{13}$ and only minimal deviation of
$\theta_{23}$ away from maximality. The other parameter, $\varepsilon_2$
leads to very tiny values of $\theta_{13}$ but to substantial deviation
of $\theta_{23}$ from maximality. Thus, the values of $\varepsilon_1$ and $\varepsilon_2$ are
determined by the magnitude of $\theta_{13}$ and the deviation of $\theta_{23}$
from maximality respectively. In the case of $\mu\leftrightarrow\tau$ symmetry,
we find that six parameters of neutrino mass matrix are needed to predict the
five neutrino oscillation parameters and the lowest neutrino mass. On the other hand,
 it is possible to obtain viable values
for the three neutrino masses and three mixing angles in terms of five
parameters by imposing $\mu\leftrightarrow-\tau$
exchange symmetry for the case of IH. However, a fine-tuned cancellation among these parameters is required
to obtain the measured value $\theta_{12}$.

\section*{Acknowledgment}
Rambabu thanks CSIR, Govt. of India and IRCC, IIT Bombay for financial support
during the course of this work. We thank Arpit Agrawal and Anindita Maiti for
various discussions.

\appendix*
 \begin{center}
 \section*{Appendix}
\end{center}
In this appendix we discuss the details of the diagonalization of
$M_4$, given in eq.~(\ref{IHb=0pert}).
First diagonalizing the $2-3$ sector
with $\theta_{23}=-\pi/4+\delta\theta_{23}$ gives
\begin{eqnarray}\label{mu-tau23}
U^T_{23}M_4U_{23}&=&
\left( \begin{array}{ccc}
1     &     0         &   0                \\
0     &     c_{23}    &   -s_{23}          \\
0     &     s_{23}    &   c_{23}         \end{array} \right)
\left( \begin{array}{ccc}
a             & -\varepsilon_1    &   \varepsilon_1     \\
-\varepsilon_1   & c-\varepsilon_2   &   d              \\
\varepsilon_1    & d              &   c+\varepsilon_2   \end{array} \right)
\left( \begin{array}{ccc}
1     &     0         &   0                \\
0     &     c_{23}    &   s_{23}           \\
0     &     -s_{23}   &   c_{23}         \end{array} \right)        \nonumber   \\
&=&\left( \begin{array}{ccc}
a                                           &   -\sqrt{2}\varepsilon_1\sin\delta\theta_{23}     &  \sqrt{2}\varepsilon_1\cos\delta\theta_{23}         \\
-\sqrt{2}\varepsilon_1\sin\delta\theta_{23}    &   c+d\sqrt{1+\frac{\varepsilon_2^2}{d^2}}         &  0                                         \\
\sqrt{2}\varepsilon_1\cos\delta\theta_{23}     &    0                                           &  c-d\sqrt{1+\frac{\varepsilon_2^2}{d^2}} \end{array} \right).
\end{eqnarray}
Here, $\tan2\delta\theta_{23}=-\varepsilon_2/d$.
After this 2-3 diagonalization, we further diagonalize the mass
matrix simultaneously 
in the 1-3 and 1-2 sectors. The form of the corresponding diagonalizing matrix for
the same is
\begin{equation}
U_{13}U_{12}=\left( \begin{array}{ccc}
c_{13}   &     0    &   s_{13}     \\
0        &     1    &   0          \\
-s_{13}  &     0    &   c_{13}   \end{array} \right)
\left( \begin{array}{ccc}
c_{12}   &  s_{12}  &   0          \\
-c_{12}  &  c_{12}  &   0          \\
0        &     0    &   1        \end{array} \right)
=\left( \begin{array}{ccc}
c_{13}c_{12}   &    c_{13}s_{12}       &   s_{13}      \\
-s_{12}        &    c_{12}             &   0           \\
-s_{13}c_{12}  &    -s_{13}s_{12}      &   c_{13}     \end{array} \right).
\end{equation}
Applying the similarity transformation with $U_{13}U_{12}$ to
$U^T_{23}M_4U_{23}$, we get
\begin{eqnarray}\label{IHspecialSimult}
(U_{13}U_{12})^TU^T_{23}M_4U_{23}(U_{13}U_{12})   
&=&\left( \begin{array}{ccc}
c_{13}c_{12}     &    -s_{12}    &   -s_{13}c_{12}    \\
c_{13}s_{12}     &    c_{12}     &   -s_{13}s_{12}    \\
s_{13}           &    0          &   c_{13}         \end{array} \right)
\left( \begin{array}{ccc}
a        & \alpha    &   \beta       \\
\alpha   & c+d'      &   0           \\
\beta    & 0         &   c-d'     \end{array} \right)    \nonumber    \\
&& \times\left( \begin{array}{ccc}
c_{13}c_{12}   &    c_{13}s_{12}       &   s_{13}      \\
-s_{12}        &    c_{12}             &   0           \\
-s_{13}c_{12}  &    -s_{13}s_{12}      &   c_{13}     \end{array} \right)
\end{eqnarray}
where $\alpha=-\sqrt{2}\varepsilon_1\sin\delta\theta_{23}, 
\beta=\sqrt{2}\varepsilon_1\cos\delta\theta_{23}$ and $d'=d\sqrt{1+\varepsilon_2^2/d^2}$.
We work out the $13$ and $23$ elements of the above matrix and set them
to be zero. We obtain the following equations
\begin{eqnarray}
 c_{13}c_{12}(as_{13}+\beta c_{13})-\alpha s_{13}s_{12}-s_{13}c_{12}[\beta s_{13}+(c-d')c_{13}] &=&  \nonumber   \\
\frac{1}{2}(a-c+d')\sin2\theta_{13}c_{12}+\beta \cos2\theta_{13}c_{12}-\alpha s_{13}s_{12}&=&0, \label{append13}  \\
 c_{13}s_{12}(as_{13}+\beta c_{13})+\alpha s_{13}c_{12}-s_{13}s_{12}[\beta s_{13}+(c-d')c_{13}] &=&   \nonumber   \\
\frac{1}{2}(a-c+d')\sin2\theta_{13}s_{12}+\beta \cos2\theta_{13}s_{12}+\alpha s_{13}c_{12} &=&0. \label{append23}
\end{eqnarray}
In the above two equations, the terms $\alpha s_{13}s_{12}$ and $\alpha s_{13}c_{12}$
can be neglected because they are of the order $10^{-6}$. With this
approximation we obtain 
\begin{equation}\label{IHspecialtheta13}
 \tan2\theta_{13}=\frac{2\sqrt{2}\varepsilon_1}{c-d'-a}\cos\delta\theta_{23}\approx-\frac{2\sqrt{2}\varepsilon_1}{a}.
\end{equation}
Diagonalization requires element $12$ also to be zero. This leads to
\begin{align}
 c_{13}c_{12}[ac_{13}s_{12}+\alpha c_{12}-\beta s_{13}s_{12}]-s_{12}[\alpha c_{13}s_{12}+(c+d')c_{12}]
-s_{13}c_{12}[\beta c_{13}s_{12}-(c-d')s_{13}s_{12}] &=& \nonumber   \\
\frac{1}{2}\sin2\theta_{12}[ac^2_{13}-\beta\sin2\theta_{13}-(c+d')+(c-d')s^2_{13}]+\alpha c_{13}\cos2\theta_{12} &=& \nonumber  \\
\frac{1}{2}\sin2\theta_{12}[(a-c-d')+(\varepsilon_1^2/a)[-2+4+2(c-d')/a]+\alpha c_{13}\cos2\theta_{12} &=&0. 
\end{align}
In the above equation we can neglect $(c-d')/a\ll 1$ and obtain
\begin{equation}
 \tan2\theta_{12}\approx-\frac{a}{d}\frac{\sqrt{2}\varepsilon_1\varepsilon_2}{a(a-c-d')+2\varepsilon_1^2}.
\end{equation}

\bibliographystyle{apsrev}
\bibliography{reference}

\end{document}